\documentclass[lettersize,journal]{IEEEtran}
\usepackage{amsmath,amsfonts}
\usepackage{array}
\usepackage{textcomp}
\usepackage{stfloats}
\usepackage{url}
\usepackage{verbatim}
\def\BibTeX{{\rm B\kern-.05em{\sc i\kern-.025em b}\kern-.08em
    T\kern-.1667em\lower.7ex\hbox{E}\kern-.125emX}}
\usepackage{balance}
\usepackage[pdftex]{graphicx}
\usepackage{subcaption}
\usepackage{cite}
\newcommand{\PM}{AutoDiCE}
\usepackage{tabularx,booktabs}
\newcolumntype{C}{>{\centering\arraybackslash}X} % centered version of "X" type
\usepackage{multirow}
\usepackage{hyperref}       % hyperlinks
\usepackage{url}

\begin{document}
\title{\PM{}: Fully \underline{Auto}mated \underline{Di}stributed \underline{C}NN Inference at the \underline{E}dge}

% \author{Xiaotian Guo\inst{1,2} \and
% Andy D.Pimentel\inst{1} 
% \and Todor Stefanov\inst{2}
% }
% %
% \authorrunning{Guo et al.}
% \institute{University of Amsterdam, The Netherlands,
% \email{\{x.guo3, a.d.pimentel\}@uva.nl}\\
% \and
% Leiden University, The Netherlands,
% \email{t.p.stefanov@liacs.leidenuniv.nl}\\
%\thanks{Use footnote for providing further
% 		information about author (webpage, alternative
% 		address)---\emph{not} for acknowledging funding agencies.} 

\author{ 

\begin{tabular}{ccc}
\href{https://orcid.org/0000-0003-4540-9013}{\includegraphics[scale=0.06]{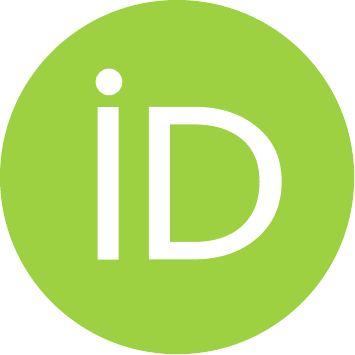}\hspace{1mm}Xiaotian Guo} & \href{https://orcid.org/0000-0002-2043-4469}{\includegraphics[scale=0.06]{orcid.pdf}\hspace{1mm}Andy D.Pimentel} & \href{https://orcid.org/0000-0001-6006-9366}{\includegraphics[scale=0.06]{orcid.pdf}\hspace{1mm}Todor Stefanov} \\
% Informatics Institute          & Informatics Institute             & Leiden Institute of \\
%  & & Advanced Computer Science         \\
University of Amsterdam, Leiden University          & University of Amsterdam            & Leiden University         \\
x.guo3@uva.nl       & a.d.pimentel@uva.nl          & t.p.stefanov@liacs.leidenuniv.nl         
\end{tabular}
}

% \author{
% Xiaotian Guo\inst{1,2} \and
% Andy D.Pimentel\inst{1} 
% \and Todor Stefanov\inst{2}
% \institute{University of Amsterdam, The Netherlands,
% \email{\{x.guo3, a.d.pimentel\}@uva.nl}\\
% \and
% Leiden University, The Netherlands,
% \email{t.p.stefanov@liacs.leidenuniv.nl}\\
% }

\maketitle

\begin{abstract}
%Abstract by Andy
Deep Learning approaches based on Convolutional Neural Networks (CNNs) are extensively utilized and very successful in a wide range of application areas, including image classification and speech recognition. For the execution of trained CNNs, i.e. model inference, we nowadays witness a shift from the Cloud to the Edge. Unfortunately, deploying and inferring large, compute- and memory-intensive CNNs on edge devices  is challenging because these devices typically have limited power budgets and compute/memory resources. One approach to address this challenge is to leverage all available resources across multiple edge devices to deploy and execute a large CNN by properly partitioning the CNN and running each CNN-partition on a separate edge device. Although such distribution, deployment, and execution of large CNNs on multiple edge devices is a desirable and beneficial approach, there currently does not exist a design and programming framework that takes a trained CNN model, together with a CNN partitioning specification, and \emph {fully automates} the CNN model splitting and deployment on multiple edge devices to facilitate distributed CNN inference at the Edge. Therefore, in this paper, we propose a novel framework, called \emph{\PM{}}, for automated splitting of a CNN model into a set of sub-models and automated code generation for distributed and collaborative execution of these sub-models on multiple, possibly heterogeneous, edge devices, while supporting the exploitation of parallelism {\em among} and {\em within} the edge devices. Our experimental results show that \PM{} can deliver distributed CNN inference with reduced energy consumption and memory usage per edge device, and improved overall system throughput at the same time.
\end{abstract}

%\begin{IEEEkeywords} edge computing; deep leanring; distributed computing; \end{IEEEkeywords}

% \fi
\IEEEpeerreviewmaketitle

\section{Introduction}
%Explain that CNNs are widely used in AI applications and why (max 2 sentences);
Deep learning (DL) \cite{lecun2015deep} has become a popular method in AI-based applications in various fields including computer vision, natural language processing, automotive, and many more. Especially, DL approaches based on convolutional neural networks (CNNs)~\cite{pouyanfar2018survey} have been extensively utilized because of their huge success in image classification~\cite{krizhevsky2012imagenet} and speech recognition applications~\cite{deng2013recent}. 

% Explain that the execution of a CNN includes 2 phases (training and inference) and describe what is done in each phase (max 3 sentences);
The execution of a CNN typically includes two phases: training and inference. During the training phase the optimal CNN parameters (i.e., weights and biases) are established. During the inference phase, a trained CNN is applied to the actual data and performs the task for which the CNN is designed. 
%State that due to the high complexity of state-of-the-art CNNs, their training and inference phases are usually performed by high-performance platforms (especially the training) and the inference is provided as cloud services by sending raw data for processing to the cloud. Explain the disadvantages of this cloud-based inference (max 4 sentences);
Due to the high complexity of state-of-the-art CNNs, the training phase is performed mainly on high-performance platforms, while the inference phase is usually provided as a cloud service, allowing less powerful compute devices at the Edge to use such services. Utilizing CNN inference as cloud services requires an edge device to send a substantial amount of data to a cloud server. Subsequently, the cloud server processes the data through the CNN and returns back the CNN result to the device. Such data communication and cloud-based computation increases the risk of data leakage from the edge device and, additionally, may cause low device responsiveness due to data transmission delays or temporal unavailability of the cloud service~\cite{cloud2010}.
% State that the above disadvantages are not tolerable by some AI applications which motivates the shift to CNN inference close to the data source, i.e., inference at the Edge.  
Evidently, this is highly undesirable for those CNN-based applications that are particularly sensitive to compute response delays or privacy of the processed data. For example, CNN-based navigation in self-driving cars \cite{dl2019patel} cannot tolerate variable and large response delays occurring due to the communication between the car and a cloud server. These delays can lead to incorrect navigation of the car and, subsequently, endanger the life of passengers. Another example is applications in medicine \cite{zhou2018unet} that use CNNs on edge devices to analyse private data of patients. Such applications cannot send their data to the cloud because this could lead to leakages of private data and violation of patients’ privacy rights. The aforementioned concerns motivate the shift of the CNN inference from the Cloud to the Edge. When entirely executed at the Edge, a CNN is deployed close to the source of data and data communication with a cloud server is not required, thereby ensuring high application responsiveness and reducing the risk of private data leakage.

%Such inference poses challenges due to the high-complexity of CNNs and the limited resources of edge devices (max 2-3 sentences).Very briefly introduce the 3 state-of-the-art approaches to address these challenges and their pros and cons. The approaches are CNN compression to fit it on a single edge device, CNN distribution along the edge-cloud continuum, and CNN distribution along multiple edge devices. Motivate why we focus on the latter approach;
Unfortunately, deploying and inferring a large CNN, which is typically memory/power-hungry and compute-intensive, on an edge device is challenging because many edge devices have limited power budgets and compute and memory resources. One approach to address this challenge is to construct a lightweight CNN model from a large CNN model by utilizing model compression techniques (e.g., pruning~\cite{reed1993pruning}, quantization~\cite{guo2018survey}, knowledge distillation~\cite{hinton2015distilling}), thereby reducing the CNN model size to a degree that allows the CNN to be deployed and efficiently executed on a resource-constrained edge device. However, the accuracy of the compressed CNN model is significantly decreased if high compression rates are required. Another approach is to infer only part of a large CNN model on the edge device and the rest on the cloud by efficiently partitioning the model and distributing the partitions {\em vertically} along the edge-cloud continuum~\cite{kang_neurosurgeon_2017}. However, the aforementioned edge device responsiveness and private data leakage issues are still inevitable in such partitioned CNN inference due to the partial involvement of the cloud. Finally, an alternative approach to address the challenge is to leverage all available resources {\em horizontally} along multiple edge devices to deploy and execute a large CNN by properly partitioning the CNN model and running each CNN partition on a separate edge device. The size of each CNN partition should match the limited energy, memory, and compute resources of the edge device the partition runs on. Such an approach not only makes it possible to deploy large CNN models without the need of model compression, respectively without loss of accuracy, but it also resolves the aforementioned responsiveness and privacy issues because a cloud server is not involved in the CNN inference. Thus, in this paper, we focus on this alternative approach, i.e., entirely distributing and executing a large CNN model at the Edge.
 
%Argue that although CNN distribution along multiple edge devices is desirable and beneficial approach, currently there is no framework available which takes a CNN together with a CNN partitioning specification and {\bf automates} the CNN model splitting and deployment on multiple edge devices in order to facilitates such approach. Therefore, in this paper, we propose a flexible framework for automated splitting of a CNN model into a set of sub-models and automated code generation for distributed execution of these sub-models on multiple edge devices, potentially exploiting parallelism among and within edge devices.
Although distributing, deploying, and executing a large CNN model on multiple, possibly heterogeneous, edge devices is a desirable and beneficial approach, currently, it requires a significant manual design and programming effort involving advanced skills in CNN model design, embedded systems and programming, and parallel programming for (heterogeneous) distributed systems. More specifically, at this moment, no design and programming framework exists that takes a trained CNN model, together with a CNN partitioning specification, and {\bf fully automates} the CNN model splitting and deployment on multiple edge devices in order to facilitate distributed CNN inference at the Edge. Therefore, in this paper, we propose a flexible framework, called {\bf\PM{}}, for automated splitting of a CNN model into a set of sub-models and automated code generation for distributed and collaborative execution of these sub-models on multiple (possibly heterogeneous) edge devices, while supporting the exploitation of parallelism {\em among} and {\em within} the edge devices.
%% \item Explain that to the best of our knowledge this is the first framework that offers the following features: here briefly highlight some important features such as flexible and easy to specify alternative CNN partitioning configurations, automated vertical CNN partitioning,
 To the best of our knowledge this is the first framework that offers the following features: 
\begin{itemize}
\item a unified interface for specifying a CNN model with Open Neural Network Exchange (ONNX) support~\cite{bai2019}, the model partitioning, and the target edge devices;
\item easy and flexible changing of the CNN model partitioning as well as mapping of partitions onto resources of edge devices;
\item automated code generation to adapt to user changes, targeting heterogeneous edge platforms; 
\item hybrid OpenMP and MPI code generation to support the exploitation of parallelism among and within the edge devices (i.e., exploiting multi-core execution);
\item a cross-platform inference engine library that supports GPU acceleration via, e.g., VULKAN and CUDA APIs. 
%ARM big.LITTLE cpu scheduling optimizations, 8-bit quantization and half-precision floating point calculation.
\end{itemize}
Our \PM{} framework is open-source and will be made available to the public at~\cite{publiccode}.

The remainder of the paper is organized as follows. Section~\ref{sec:relatedwork} discusses related work, after which Section~\ref{sec:theframework} presents our \PM{} framework. Section~\ref{sec:framworkevalution} describes a range of experiments, demonstrating that \PM{} can easily and rapidly realize a wide variety of distributed CNN inference implementations with diverse trade-offs regarding energy consumption, memory usage and system throughput. Section~\ref{sec:Discussion} provides a short discussion on the current version of \PM{} and how it could be further improved in the future. Finally, Section~\ref{sec:conclusion} concludes the paper.

\section{Related Work}
\label{sec:relatedwork}
% *****************
% % samples template. training; privacy-preserving. vs inference.
% tricks for parallel distributed training gpipe xpipe.
% are they also support automated partitioning while training.  ---> model parallelism. how automatically it could achieve.
% give an example of complicated cnn
% focus on cnn.
Today's convolutional neural network (CNN) models for computer vision tasks are becoming increasingly complex.
%, like residual blocks with a very deep structure\cite{he2016deep}, long short-term memory (LSTM) \cite{sak2014long} and transformer\cite{attention2017}, etc. 
For example, the CNN-based model CoAtNet-7~\cite{dai2021coatnet} reaching the highest top-1 accuracy of $90.88\%$ for the ImageNet dataset has 2.44 billion parameters (weights and biases) which values have to be determined during the training and stored/used during the inference. To train and deploy such large CNN models, parallel or distributed computing is often required. 
For model training, a common approach to accelerate the training process is to exploit pipeline parallelism.
%there are two main approaches to parallelize the training process across multiple GPUs \cite{pipeDream2019}, i.e., approaches exploiting data parallelism and/or model parallelism.
For example, GPipe~\cite{huang2019gpipe} applies pipeline parallelism by splitting a mini-batch of training data into smaller micro-batches, where different GPUs train on different micro-batches. Another example is PipeDream~\cite{pipeDream2019} which partitions the CNN model for multiple GPUs such that each GPU trains a different part of the model. An alternative distributed training approach, motivated by privacy concerns among multiple devices/machines, is federated learning (FL) \cite{Limsurvey2020, comprehensiveflyin2021}.
%, proposed in \cite{konevcny2016federated,kairouz2019advances, wang2019adaptive,li2020federated, savazzi_framework_2021} and reviewed in survey~\cite{Limsurvey2020, comprehensiveflyin2021}, 
FL aims at training a global centralized model with multiple, local datasets on distributed devices or data centers, thereby preserving local data privacy and improving learning efficiency. 
%Such existing FL methods are characterized with preserving data privacy and improving learning efficiency. 
%Beyond that, model parallelism is widely used in distributed training for DNN models that are too large to train in a single GPU or machine. 
%For example, PipeDream\cite{pipeDream2019} partitions the DNN model into multiple GPUs such that each GPU trains different part of the model. PipeDream achieves 5$\times$ faster than data-parallel training by pipelining the DNN training execution of each GPU. 
All of the aforementioned approaches target efficient, distributed training of large CNN models. In contrast, our work presented in this paper focuses on efficient, distributed inference of large CNNs.
%for single-batch input data.
%especially on improving training efficiency (time-to-accuracy) by overlapping computation and communication on distributed GPUs or machines. 
%But our method focuses on the distributed inference of DNNs for single-batch input data.

% | goal difference.
%%%%% introduction text unnecessary
% have limitations on CNN model support.  platform architectures, operations of CNN models (fully connected or convolutional layers), , etc. .  
%%%%%%%%%%%%%%%%
% keep 2 or 3 important references.
% reduce related work references.

% The FL researches are characterized with preserving privacy, improving learning efficiency and defending against attacks and failures.
Unlike the parallel or distributed CNN training, discussed above, the inference of large CNN models often needs to take multiple requirements into account, such as latency, throughput, resource usage, power/energy consumption, etc. To satisfy these requirements when executing the inference of large CNNs on edge devices, the following two approaches for distributed CNN model inference are typically used: {\em vertically} and {\em horizontally} distributed inference. 

In {\em vertically} distributed inference (e.g., \cite{ kang_neurosurgeon_2017,teerapittayanon_distributed_2017, li_edge_2018}), the workload of a large CNN is distributed along the cloud-edge continuum. Such an approach maximizes the utilization of computing resources on edge devices, reduces the computation workload on the cloud, and usually improves the CNN inference throughput. 
The most common idea in this approach is to obtain a specific small sub-model from or an early-exit branch of the initial large CNN model that runs on the edge device. Only if the inference result of the deployed sub-model/early-exit branch on the edge device is below a certain confidence threshold, the device has to upload its data on the cloud and the CNN inference has to continue on the cloud.
%For example, NeuroSurgeon \cite{kang_neurosurgeon_2017} partitions the origin model into two parts and deploys its small part on an edge device. 
%Other method, device-edge synergy \cite{teerapittayanon_distributed_2017,li_edge_2018,huang_deepadapter_2020} jointly retrains the extended new early-exit branches based on origin model. 
Vertical distribution along the cloud-edge continuum still relies on the quality and stability of network connections between the edge device and the cloud server because intermediate results of the small CNN sub-models or early-exit branches may still need to be uploaded to the cloud. This not only suffers from high communication latency but also there is a risk of information leakage. In contrast, our framework achieves lower inference latency by deploying a large CNN model over edge devices without the cloud, and therefore also preserves both data and model privacy.  

In {\em horizontally} distributed inference (e.g., \cite{mao_modnn_2017,zhao_deepthings_2018,stahl_fully_2019, hadidi_toward_2020}), the workload of a large CNN is fully distributed among multiple edge devices. That is, all CNN computations are collaboratively executed at the Edge and there is no dependency on the cloud. Data partitioning and model partitioning are two common methods to horizontally distribute the CNN inference across multiple edge devices. 
Data partitioning exploits data parallelism among multiple devices by splitting the input/output data to/from CNN layers into several parts while each device executes all layers of a CNN model using only some parts of the data. For example, DeepThings \cite{zhao_deepthings_2018} uses the Fused Tile Partitioning (FTP) method for splitting input data frames of CNN layers in a grid fashion to reduce the CNN memory usage. 
%Each partitioned part of the input data forms independently distributive task that is scheduled in DeepThings runtime system. 
The main drawback of the data partitioning method is that an edge device should still be capable of executing all layers of a CNN model which implies that the edge device should be able to store the weights and biases of the entire CNN model.
% still store all weights.
% related work, cnn ---> dnn.
Alternatively, the model partitioning method splits the CNN layers and/or connections of a large CNN model, thereby creating several smaller sub-models (model partitions) where each sub-model is executed on a different edge device~\cite{stahl_fully_2019}. 
%Alternatively, each device could also execute a different part (i.e., partition) of the origin CNN model \cite{stahl_fully_2019,hadidi_musical_2018}. CNN models can be partitioned into sub-models horizontally by splitting layers, or vertically by splitting layer connections. 
For example, MoDNN~\cite{mao_modnn_2017} splits convolution layers and fully connect layers in the VGG-16 model. In~\cite{hadidi_toward_2020}, CNN layer connections are split and each CNN layer is treated as a sub-task. These sub-tasks are then mapped to edge devices through a balanced processing pipeline approach.

The aforementioned efforts for horizontally distributed inference focus on performance optimization through partitioning, scheduling and exploiting parallelism, whereas our work is complementary to these efforts as it targets the actual automation of partitioning, code generation, and model deployment for distributed CNN inference at the Edge. Our framework is flexible for users to easily explore objectives of distributed CNN inference at the Edge such as reducing memory usage and energy consumption per edge device, improving CNN inference latency/throughput, etc. 

%In addition, these efforts  often have limitations with respect to automated deployment due to third-party dependencies. For example, the work presented in~\cite{hadidi_toward_2020} uses the Keras $2.1$ framework with a TensorFlow back-end and other framework dependencies. Deep learning frameworks such as Keras, TensorFlow, Pytorch, etc. are designed for training of CNN models with sufficient resources, and are less suitable for targeting multiple, possibly heterogeneous, resource constrained devices. In contrast, our C++ based inference engine is designed for fast and efficient deployment of CNN models on heterogeneous devices. Moreover, the automation deployment is complemented with SPMD code generation which makes it more convenient over multiple devices. 
 
\section{The \PM{} framework}
\label{sec:theframework}
In this section, we describe our \PM{} framework as a design flow and explain the main steps in the flow with the help of an illustrative example. First, we provide a high-level overview of the \PM{} design flow. Second, we describe \PM{}'s unified user interface. Next, we explain in detail the main steps in the front-end of the \PM{} design flow. Finally, we do the same for the back-end of the flow.

\subsection{Overview}
\label{sec:overview}
%Briefly introduce \PM, which is a unified framework for distributed CNN inference at the edge. (max 2 sentences); 
\PM{} is a flexible framework that facilitates distributed inference of a CNN model, embedded in an AI application, at the Edge. More specifically, it allows designers and programmers of such CNN-based AI applications to perform,~\emph{in a fully automated manner}, CNN model partitioning, deployment and execution on multiple resource-constrained edge devices. 
\begin{figure}[!t]
 \centerline{\includegraphics[width=0.5\textwidth]{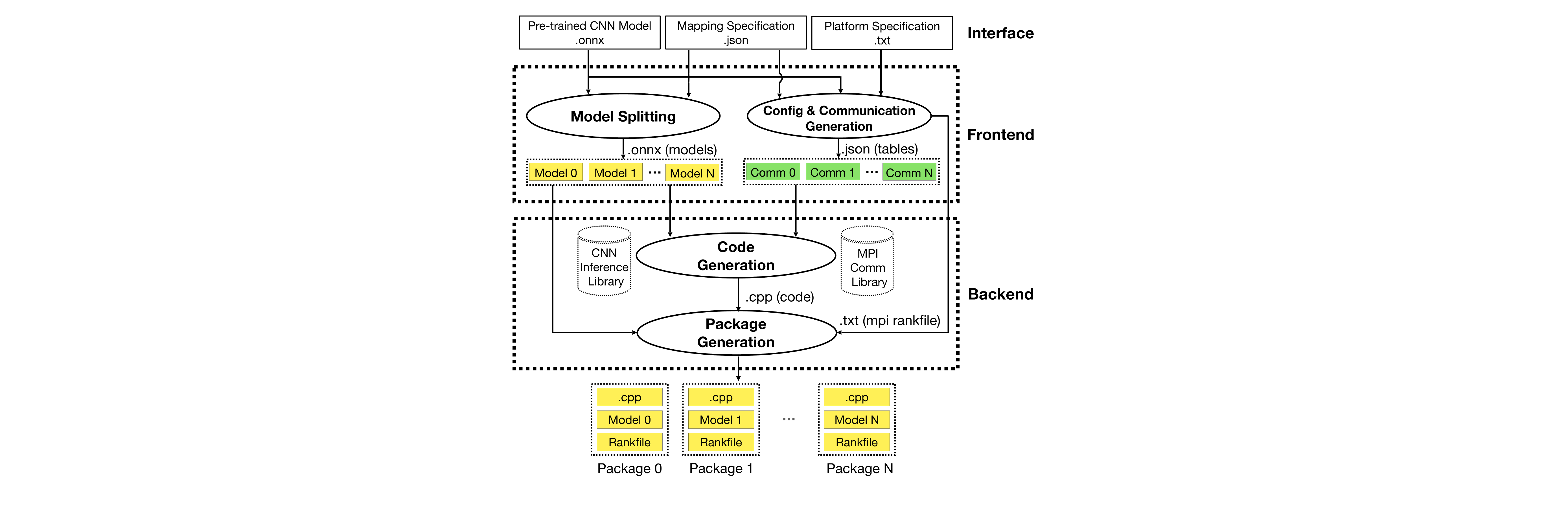}}
 \caption{The~\PM{} design flow and its user interface}
 \label{design}
\end{figure} 
 % Present the framework design in Figure \ref{designflow}. \figurename{\ref{designflow}}. State that it contains three parts, the first part provides users a unified interface to specify computing nodes, models, and layer mapping over these nodes. The last two parts, frontend and backend, are the main building components of \PM{} to deploy distributed CNN applications. 
Figure \ref{design} shows the \PM{} user interface and design flow where the main steps in the flow are divided into two modules: front-end and back-end. 

The interface is composed of three specifications, namely Pre-trained CNN Model provided as an .onnx file, Mapping Specification provided as a .json file, and Platform Specification provided as a .txt file. 
%Typically, these layers are connected in a feed-forward fashion to compose a CNN model, and each layer stores its own coefficients (weights and bias).

The Pre-trained CNN Model specification includes the CNN topology description with all layers and connections among layers as well as the weights/biases that are associated with the layers and obtained by training on a specific dataset using deep learning frameworks like PyTorch, TensorFlow, etc. Many such CNN model specifications in ONNX format~\cite{bai2019} are readily available in open-access libraries and can be directly used as an input to our framework.

The Platform Specification lists all available edge devices together with their computational hardware resources and specific software libraries associated with these resources. This specification is simple to draw up and can be generated by external tools that query the network connecting the edge devices or provided manually by the user. 

The Mapping Specification is a simple list of key-value pairs in JSON format that explicitly shows how all layers described in the Pre-trained CNN Model specification are mapped onto the computational hardware resources listed in the Platform Specification. Every unique key corresponds to an edge device with a selection of its hardware resources to be used for computation. Every value corresponds to a set of CNN layers to be deployed and executed on the edge device resources. Such a Mapping Specification can be generated by external system-level design-space exploration (DSE) tools or provided manually by the user.   

The three aforementioned specifications are given as an input to the front-end module as shown in Figure~\ref{design}. Two main steps are performed in this module:~\emph{Model Splitting} and ~\emph{Config \& Communication Generation}. The Model Splitting takes as an input the Pre-trained CNN Model and Mapping specifications, splits the input CNN model into multiple sub-models, and generates these sub-models in ONNX format. The number of generated sub-models is equal to the number of unique key-value pairs in the Mapping Specification. Each sub-model contains input buffers, output buffers, and the set of CNN layers, specified in the corresponding key-value pair. The Config \& Communication Generation step takes all three specification files as an input and generates specific tables in JSON format containing information needed to realize proper communication and synchronization among the sub-models using the well-known MPI interface. In addition, a configuration text file (MPI rankfile) is generated to initialize and run the sub-models as different MPI processes. %As we know, MPI program has multiple MPI processes, and each MPI process has its own rank number for identification. This file maps individual MPI processes to multiple computing nodes, and also assigns each computing node with a rank number. 

As shown in Figure~\ref{design}, the generated configuration file, sub-models, and tables are used in the back-end module for code and deployment package generation. During the~\emph{Code Generation} step in this module, efficient C++ code is generated for every edge device based on the input sub-models and tables. In the generated code, primitives from the standard MPI library are used for data communication and synchronization among sub-models as well as primitives from our customized CNN Inference Library are used for implementation of the CNN layers belonging to every sub-model. Both libraries enable the generation of cross-platform code that can be compiled for and executed on multiple heterogeneous edge devices. Finally, the~\emph{Package Generation} step packs the generated cross-platform C++ code, the MPI rankfile, and a sub-model together to generate a specific deployment package for every edge device. All packages contain the same C++ code and the same MPI rankfile but different sub-models. When a package is compiled, deployed, and executed on an edge device, the specific sub-model in the package will be loaded and only the part of the code that corresponds to the loaded sub-model will run as an MPI process as specified in the MPI configuration rankfile. 
%It is well worthwhile to provide an easy-to-use interface for users to execute distributed inference. Such that our framework can easily and quickly generate actual implementations to distribute a CNN model at the edge.

\begin{figure*}[!t]
\centering
  \includegraphics[width=\textwidth]{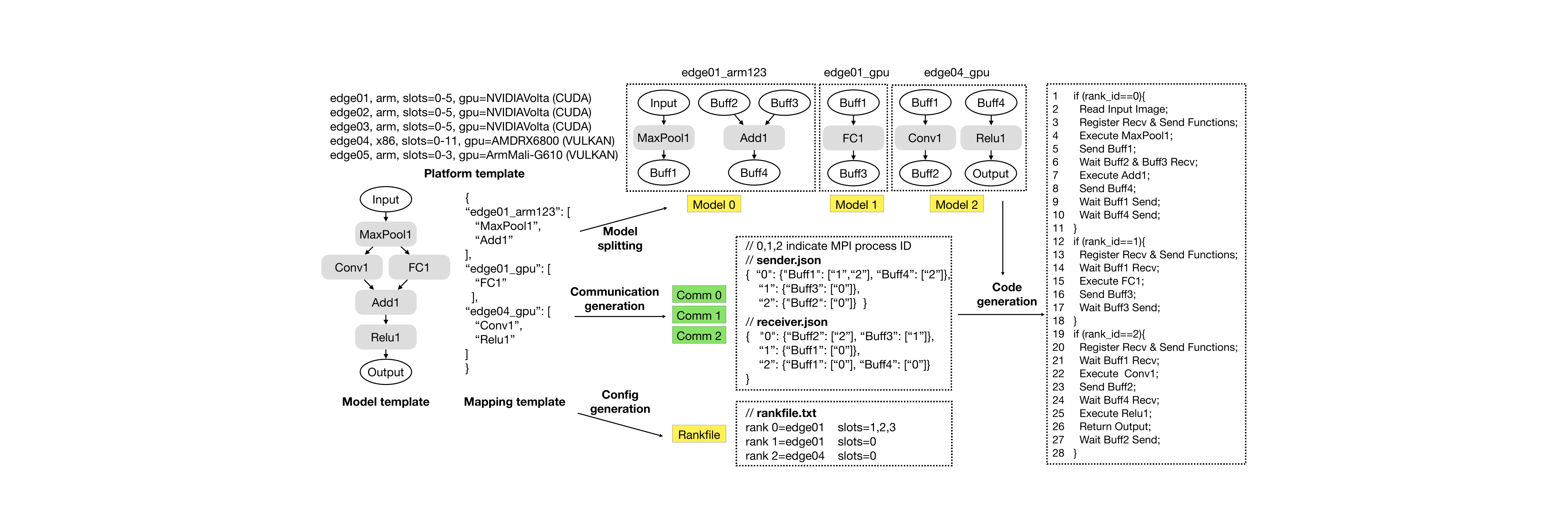}
  \caption{\PM{} in action: a detailed example}
  \label{detailframework}
\end{figure*} 

In the following subsections, the interface and the main steps of the \PM{} design flow, introduced above, are explained in more detail with the example in Figure~\ref{detailframework}.

\subsection{Interface}
\label{sec:Interface}
 %illustrated as the three templates in the left-most part of Figure 2.
In the left-most part of Figure~\ref{detailframework}, we show three templates (examples) representing the three specifications of the user interface introduced in Section~\ref{sec:overview}. By using these example templates, we comprehensively reveal and explain the flexibility of and heterogeneity support in~\PM{}.
  
% The platform specification provides explicit information of the target hardware platform(s), such as name of each computing node, cpu/gpu architecture, available cpu/gpu resources, etc. The Platform Specification lists all available edge devices together with their computational hardware resources and specific software libraries associated with these resources. This specification is simple to draw up and can be generated by external tools that query the network connecting the edge devices or provided manually by the user. 
In general, the Platform Specification lists all available edge devices with their computational resources. Every line in the list specifies the name of the edge device, the CPU architecture, the number of CPU cores, and (optionally) a GPU device with its architecture and programming library. For instance, the first line of the platform template in Figure~\ref{detailframework} specifies that the name of the device is~\textit{"edge01"} with an ARM processor architecture including six cores in total (slots=0-5) and one GPU device with NVIDIAVolta architecture supported by the CUDA library.
% explains the heterogeneity of platforms like different cpu architectures such as arm, x86/64, etc., gpu architectures such as Nvidia, Mali etc., different GPU programming APIs such as CUDA, Vulkan, etc. Explain the example in Figure 2.
Through the Platform Specification, a user can easily and flexibly specify alternative heterogeneous hardware platforms including different numbers of edge devices and type of resources. As shown in Figure~\ref{detailframework}, the user can select different CPU architectures per edge device such as ARM, x86, etc. with different numbers of cores as well as  different GPU architectures per edge device such as NVIDIA, Mali, AMDRX, etc. with different GPU programming APIs such as CUDA, VULKAN, etc. 

% Briefly introduces what is a CNN model: consists of layers, and coefficients of each layers. 
The model template in~Figure~\ref{detailframework} is an example of a part of a Pre-trained CNN Model specification that visualizes the CNN model topology only. It contains an input layer, five hidden layers (i.e., MaxPool1, Conv1, FC1, Add1, and Relu1), and an output layer. Every hidden layer stores its own parameters (such as weights, bias, etc.) that are not shown in Figure~\ref{detailframework}.
%states why \PM{} chooses standard ONNX model as input.
In order to support interoperability of \PM{} with other DL frameworks, we adopt ONNX as the standard format to represent/specify a pre-trained CNN model in the \PM{} interface. The choice of ONNX allows users to provide a CNN model designed, trained, and verified in well-known and widely-used frameworks such as TensorFlow~\cite{tensorflow2015-whitepaper}, PyTorch~\cite{pytorch2019}, etc. A large variety of trained CNN models are already available in ONNX format that can be readily utilized by~\PM{}, allowing easy deployment of these models over multiple edge devices. 
In addition, the use of ONNX facilitates reproducibility in terms of CNN designs (e.g., CNN topology, used parameters, etc.) and in CNN evaluations (for CNN model accuracy and non-functional characteristics). For example, in experimental evaluations, users can confidently and reliably compare CNN model characteristics such as accuracy, memory usage, performance, and power/energy consumption, obtained by~\PM{}, with the same characteristics obtained by other frameworks and approaches, applied on exactly the same CNNs. 

%introduces  how  the  mapping  template  works.  A  mapping file  is  a  dictionary  which  contains  keys  and  values.Keys  represent  the  deployed  computing  nodes  (from the platform specification), whereas the values contain the  CNN  model  layer  names  that  will  be  executed  on a  particular  node.  A  computing  node  indicates  which device is used and how many cpu cores or gpu are used in the device.

As mentioned in Section~\ref{sec:overview}, the Mapping specification lists several different key-value pairs to describe a distribution of the layers in a CNN model over different computational platform resources. The Mapping template in Figure~\ref{detailframework} is an example of such specification. It lists three different key-value pairs. For example, the unique key \textit{"edge01\_arm123"} specifies that three ARM CPU cores (i.e., cores 1, 2, and 3) of device \textit{edge01}, described in the Platform specification, are allocated for CNN layers execution. The corresponding value [\textit{"MaxPool1", "Add1"}] specifies that layers MaxPool1 and Add1, described in the Pre-trained CNN model specification, are executed on the allocated three cores.
%Emphasize  the  flexibility  of  keys.  To  give  maximum flexibility, users can bind CNN layers to a single GPU,a single cpu core or multiple cpu cores. Specifically, if all keys are using the same device, the inference turns into multi-threads on a single device.
All valid keys must be generated from the Platform Specification to ensure the availability of chosen computational resources. Users can bind CNN layers to a single GPU, a single CPU core or multiple CPU cores. Specifically, if all keys use computational resources of the same device, the distributed inference turns into a multi-threaded execution on a single device. 
%Usually an edge device in a real-world IoT cluster exists resident applications such as sensor data acquisition, video surveillance, etc., not all computing resources of the edge device are available. The flexibility of key choices is fully applicable to distribute CNN-based applications in the real-world scenarios. 
%\PM{} supports different approaches for splitting (and the parallel execution of) a CNN model (vertical, horizontal and data parallelism), but that this paper will only focus on vertical splitting. This means that a CNN layer is mapped to only one unique key (device). 
All valid values must be selected from layers of the Pre-trained CNN model, and all CNN layers in that  model should be assigned to at least one hardware processing unit (CPU or GPU) to ensure the mapping consistency. The mapping example in Figure \ref{detailframework} is a vertical partitioning, which means that every CNN layer is mapped to a single unique key (device). If a CNN layer is mapped to multiple unique keys, then the layer will be horizontally distributed over multiple computational resources. Users can realize different approaches for splitting (and parallel execution of) a CNN model, namely vertical, horizontal and using data parallelism (the latter two are not shown in Figure \ref{detailframework}). This is done by changing the layer distribution in the Mapping Specification. However, in this paper, we will only focus on vertical partitioning. 
It is easy and flexible for users to change the CNN model partitioning as well as mapping of partitions to edge devices through selecting different combinations of key-value pairs in the Mapping Specification.
 
\subsection{Front-end}
\label{sec:frontend}
%The main function of the frontend is to split the model (according to the mapping specification) as well as to generate communication tables and an MPI-specific configuration file.
The front-end module is designed to parse, check, and pre-process all user specifications through its two main steps: Model Splitting and Config \& Communication Generation. Model Splitting splits the input CNN model according to the mapping specification and generates several CNN sub-models. Each sub-model will be implemented and executed as an MPI process. Config \& Communication Generation generates an MPI-specific configuration file and communication tables based on the three input specification files.
%In Figure. 2, explains what model splitting does: splits the  original  graph  into  multiple  sub-graphs  according to the mapping file. Each sub-graph forms a sub-model by adding input and output nodes.
At the top center of Figure \ref{detailframework}, the model splitting step is illustrated. Based on the three key-value pairs in the Mapping template (specification), the CNN model template is vertically partitioned into three sub-models (\textit{Model 0, Model 1, and Model 2}). The layers of the CNN model mapped on the same edge device resource will be grouped into a single sub-model. For example, the two layers \textit{MaxPool1} and \textit{Add1} are grouped together to form a sub-model \textit{Model 0}. 
% states all layers compose a directed acyclic graph(DAG),  which  each  layer  can  be  treated  as a graph node.  Attributes of a graph node includes coefficients and layer definitions. All layers compose into a directed acyclic graph in a complete CNN model. 

%states the input and output nodes in each sub-models are actually connected with each other in the original model, as the output of a CNN layer is the input of the next connected CNN layer.
The output of a CNN layer in the initial Model template is the input of its next connected CNN layers. If two connected CNN layers are mapped onto different edge devices or different compute resources (CPU or GPU) within an edge device, i.e., the two layers belong to two different sub-models, the direct connection between these two layers is replaced by one output buffer belonging to one of the sub-models and one input buffer belonging to the other sub-model. These two buffers are used to store and communicate intermediate results between the two CNN layers. For example, the directly connected CNN layers \textit{MaxPool1} and \textit{Conv1} of the Model template in Figure~\ref{detailframework} are mapped onto two different edge devices according to the Mapping template. Thus, layer \textit{MaxPool1} belongs to sub-model \textit{Model 0} and layer \textit{Conv1} belongs to sub-model \textit{Model 2}. As a consequence, the direct connection between \textit{MaxPool1} and \textit{Conv1} is replaced by output buffer \textit{Buff1} in \textit{Model 0} and input buffer \textit{Buff1} in \textit{Model 2}. 

%%  give an short introduction of rankfile, which gives users detailed control of binding MPI processes to physical units (PU). Rankfiles are text files that specify detailed information about how individual processes should be mapped to nodes, and to which processor(s) they should be bound. 
The Config Generation step is illustrated in the bottom center of Figure \ref{detailframework}. It  generates an MPI-specific Rankfile which provides detailed information about how the individual MPI processes, corresponding to the generated sub-models, should be mapped onto edge devices, and to which processor/core(s) of an edge device an MPI process should be bound to. In the example in Figure~\ref{detailframework}, we have three sub-models \textit{Model 0, Model 1, and Model 2} that will be implemented and executed as three 
different MPI processes 0, 1, and 2, respectively. Based on the Mapping template, the example Rankfile in Figure~\ref{detailframework} specifies that the MPI processes 0 and 1 should be mapped onto edge device \textit{edge01} and the MPI process 2 should be mapped onto edge device \textit{edge04}. In addition, each line of the Rankfile specifies the physical processors/cores allocated to the corresponding MPI process. In our example Rankfile, the first line 
specifies that MPI process 0 should be mapped on edge device \textit{edge01} and slots 1, 2, and 3 are allocated to this process on this device. This means that this process will run on three ARM CPU cores (i.e., core 1, 2, and 3) of device \textit{edge01}.

%If these two layers are mapping into different edge nodes, the connection between these two layers is cut off. Therefore we generate communications between different sub-models according to the mapping file. 
The Communication Generation step is illustrated in the center of Figure \ref{detailframework}. It generates a sender table and a receiver table as .json files. These two communication tables specify the necessary communications between individual MPI processes to ensure that the input/output buffers of the corresponding sub-models are synchronized through the MPI interface. For example, the first line in the sender table specifies that MPI process 0 needs to send the contents of \textit{Buff1} to MPI processes 1 and 2, and the contents of \textit{Buff4} to MPI process 2. Correspondingly, the third line in the receiver table specifies that MPI process 2 needs to receive the contents of \textit{Buff1} and \textit{Buff4}, both from MPI process 0. The communication and synchronization information in the sender and receiver tables ensure that the initial input CNN model is correctly executed after the model splitting.   

%Also stress that consecutive layers in the CNN model mapped to the same edge resource will be grouped in a single sub-model (i.e., without explicit communication taking place)

\subsection{Back-end}
%Automated code generation step is the core of back-end module. It is essential in the design as it highly improves the programming productivity instead of writing by hand. Second, it makes the program more understandable by creating a middle space between user definition and code implementation. 
The back-end module constitutes the \PM{}'s final stage to create a CNN-based application for deployment over multiple edge devices. It contains two main steps: Code Generation and Package Generation. 
% explains that automated code generation is essential in the design as it highly improves the programming productivity instead of writing by hand. Second, it makes the program more understandable by creating a middle space between user definition and code implementation.

The first step, Code Generation, turns all intermediately generated files (all sub-models and communication tables) by the front-end module into efficient C++ code. The output of this step is a single .cpp file which has a very specific and well-defined code structure, making calls to specific primitives and functions located in two libraries: a standard MPI Library and our customized CNN Inference Library. The code structure contains several code blocks. Each code block is surrounded by an {\tt if} statement and implements one CNN sub-model. The sub-models are executed as individual MPI processes mapped on different edge device resources, meaning that every MPI process runs only the code block implementing the corresponding sub-model. The code block is uniquely identified by a rank ID checked in the {\tt if} statements surrounding the code blocks.
Unique rank IDs are assigned according to the Rankfile, explained in Section~\ref{sec:frontend}, during
the MPI initialization stage. The pseudo code template in the right-most part of Figure~\ref{detailframework} illustrates the specific code structure of the generated .cpp file. It contains three code blocks, i.e., Lines 1-11, Lines 12-18, and Lines 19-28, that implement sub-models \textit{Model 0, Model 1, and Model 2}, respectively. \textit{Model 0, Model 1, and Model 2} will be executed as three MPI processes 0, 1, and 2, respectively. Every MPI process contains the aforementioned code template but the MPI process 0 corresponding to sub-model \textit{Model 0} will run only the code block between lines 1 and 11. Similarly, the MPI process 1 will run only the code block between lines 12 and 18, etc.

All code blocks have a similar, well-defined structure starting with code that registers all MPI send and receive primitives (e.g., lines 3, 13, and 20 in Figure~\ref{detailframework}) followed by MPI\_Wait primitives that block the code execution until the necessary data to be processed by CNN layers is received (e.g., lines 6, 14, 21, and 24). Then, code implementing the CNN layers is executed followed by MPI\_Send primitives that communicate the output data from a layer to other layers executing in different MPI processes mapped on different edge devices/resources (e.g., lines 7-8, 15-16, 22-23). Finally, MPI\_Wait primitives are used to block the code execution until the sent data arrives at the destination (e.g., lines 9, 10, 17, and 27).

%It highly improves the programming productivity instead of writing by hand and makes the program more understandable by creating a middle space between user definition and code implementation. 
% The code generation module takes all sub-models and communication tables as its input to generate C++ codes. Here, we need to explain that we need our own inference library to generate all CNN code, as the library has been optimized for (distributed) execution on resource-constrained edge devices. 
% illustrate the generation steps  
%More specifically, this step turns each sub-model into different code snippets (e.g., ARM NEON, VULKAN, CUDA) according to its specified target hardware platform (e.g., CPU, GPU). Moreover, it transforms the communication tables into MPI primitives for data communication between the different sub-models. The pseudo code template in the right-most part of Figure ~\ref{detailframework} illustrates the code generation step. Here, the individual MPI processes (executed on different resources) run a different part of the code according to its rank id. For example, MPI process 0 runs the first code block in the {\tt if} statement and only loads Model 0. This unique rank id is assigned according to the rankfile during the MPI initialization stage. 

Some code blocks have to implement and execute more than one CNN layer because the corresponding CNN sub-models contain multiple CNN layers. Every code block implementing multiple CNN layers has to execute the layers in the order specified by the data dependencies in the input CNN Model template to preserve the functional correctness of the distributed CNN model. For example,  the CNN sub-model \textit{Model 0} in Figure~\ref{detailframework} is implemented by the code block between lines 1 and 11 in Figure~\ref{detailframework}. Line 2 reads an image file to prepare the input data for the CNN model. The code in line 3 registers all non-blocking MPI send and receive primitive calls according to the first lines in the sender and receiver tables, explained in Section~\ref{sec:frontend}. In lines 4 and 7, the \textit{MaxPool1} and \textit{Add1} layers are executed one after the other, thereby preserving the order specified in the CNN Model template given in Figure~\ref{detailframework}. After executing each layer, they store their output data in \textit{Buff1} and \textit{Buff4}, respectively. Line 5 sends the content of \textit{Buff1} to MPI process 1 and MPI process 2 according to the sender table. The used non-blocking MPI\_Send primitive returns immediately and will not block the execution. A layer within a code block is executed once its input data is available, i.e., layers are executed in a data-driven fashion. For those layers that read their input data from communication buffers (i.e., data generated by another sub-model, possibly running on a different edge device), MPI synchronization (wait) primitives enforce that layers cannot start execution before their input data is available. For example, this data-driven based execution of layers enforces that the \textit{Add1} layer in \textit{Model 0} can only be executed after the input data in \textit{Buff2} and \textit{Buff3} is available. Such synchronization is realized by the MPI\_Wait primitives in line 6 of Figure~\ref{detailframework}.
%The MPI\_Wait primitive used in line 6 blocks the execution until the contents of \textit{Buff2} and \textit{Buff3} are received. 
Line 8 uses the non-blocking MPI\_Send primitive again to transfer the content of \textit{Buff4} to MPI process 2. Finally, at the end of the code block, in lines 9-10, two synchronization MPI\_Wait primitives are called that are associated with the two asynchronous send requests in lines 5 and 8. All such synchronization primitives are always called at the end of a code block in order to stop the code execution until the corresponding send requests (in this example the requests to send the contents of \textit{Buff1} and \textit{Buff4}) are completed.
 
 % Introduce that the execution of sub-models depends on our CNN inference engine. 
In every code block, the implementation and execution of the CNN layers is realized by calling functions and primitives located in our custom CNN Inference Library.
%The layer executions of the sub-model mainly depend on our own inference library. 
By encapsulating the NCNN~\cite{ncnn} and Darknet~\cite{darknet13} neural network frameworks into a uniform wrapper, our custom inference library supports CNN layer implementation and execution on a variety of hardware platforms (e.g., Raspberry Pi with a quad-core ARM v8 SoC, NVIDIA Jetson AGX Xavier series, etc.). 

The used MPI primitives in the code blocks are part of the Open MPI library~\cite{gabopenmpi}, which is an open-source implementation of the standard MPI interface for high performance message passing. It enables parallel execution on both homogeneous and heterogeneous platforms without drastic modifications to the device-specific code. 

\begin{figure}[!t]
 \centerline{\includegraphics[width=0.5\textwidth]{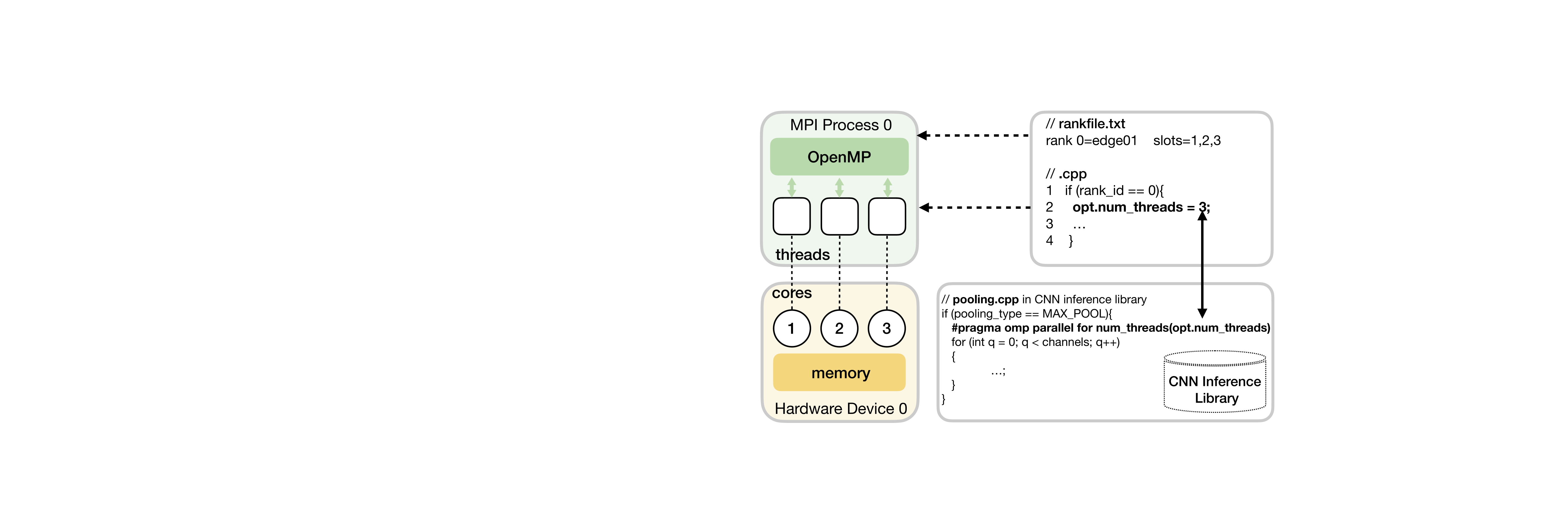}} 
 \caption{MPI process 0 with OpenMP}
 %\centerline{\includegraphics[width=0.5\textwidth]{image/mpiruntime.pdf}}
 %\caption{MPI + Open MPI hybrid programming Model}
 \label{mpiopenmp}
\end{figure}

Besides facilitating the C++ code generation and distributed execution of CNN models (using MPI), our customized CNN Inference Library also integrates and provides OpenMP support.
This means that if a CNN layer is mapped onto multiple CPU cores in an edge device, the actual execution of such layer will be multi-threaded using OpenMP in order to efficiently utilize the multiple CPU cores by exploiting data parallelism available within the layer. 
For example, the \textit{MaxPool1} layer in Figure~\ref{detailframework} is implemented and executed as multiple threads within MPI process 0 which is mapped onto the three ARM CPU cores 1, 2 and 3 in edge device \textit{edge01}. More specifically, in Figure~\ref{mpiopenmp}, we show some details about how the multiple threads bound to the three CPU cores 1, 2 and 3 are executed within MPI process 0. A thread number variable, called \textit{num\_threads}, is set to 3 in the code block implementing MPI process 0 during the code generation step. In our customized CNN Inference Library, this variable is used in the implementation code of all types of layers (i.e., convolution, pooling, etc.), and it configures the OpenMP macro line \textit{\#pragma omp parallel for} shown in Figure~\ref{mpiopenmp}. This macro line spawns a group of multiple threads and divides the loop iterations (the {\tt for} loop in Figure~\ref{mpiopenmp}) that follow this macro line between the spawned threads during the execution. So, during the execution, layer \textit{MaxPool1} is executed as three threads running on CPU cores 1, 2, and 3.

The above discussion on the first step (Code Generation) of the back-end module clearly indicates that our framework employs a hybrid MPI+OpenMP programming model. OpenMP is used for parallel execution of a CNN layer within an edge device and MPI is used for communication and synchronization among CNN sub-models running on different edge devices or on different compute resources (e.g., CPUs and GPUs) within an edge device. By doing so, our framework provides extreme flexibility in terms of many alternative ways to distribute the CNN inference within and across edge devices by treating every CPU core or GPU unit in edge devices as a separate entity with its own address space. This allows our framework to be used in very complex IoT scenarios that may contain a lot of heterogeneous devices.

The second step of the back-end module, i.e. Package Generation, packs the generated .cpp code, sub-models, and Rankfile together into a deployment package for every edge device utilized in the distributed CNN inference. As it is essential to identify the individual MPI process running on an edge device, this step must put the Rankfile in every package. The Rankfile provides detailed information about the MPI processes' binding, which constrains each MPI process to run on specific compute resources of different edge devices. The executable binary (to be deployed on an edge device) will be generated when the corresponding .cpp code in a package is compiled together with the aforementioned CNN Inference Library. As all packages contain the same .cpp code (i.e., we use the Single Program Multiple Data paradigm in this sense), the same binary can be deployed and executed on the same type of edge devices where each edge device will load the corresponding CNN sub-model from its own package before the execution of the binary. For different types of edge devices, we can generate an executable binary for every type.

\begin{table*}[t]
  \caption{Used CNN models and \PM{} execution time breakdown}
  \label{CNNmodels+ET}
  \centering
\begin{tabular}{@{}lcccccc@{{}}}
\toprule
Network & Total \# & Total \#  & Memory for & \multicolumn{3}{c}{\PM{} Execution Time (seconds)} \\ \cline{5-7}
  & Layers & Parameters & Parameters (MB) & Front-end & Back-end & Package deployment \\
\midrule
\midrule
DenseNet-121~\cite{huang2019convolutional} & 910 & 8.06 million & 32 & 1.93 & 0.3 & 21.3\\
\midrule
ResNet-101~\cite{He2015} & 344 & 44.6 million & 171 & 7.30 & 0.1 & 23.3 \\
\midrule
VGG-19~\cite{Simonyan15} & 47 & 143 million & 549 & 21.50 & 0.4 & 26.9 \\
\midrule
\end{tabular}
\end{table*}

\section{Framework Evaluation}
\label{sec:framworkevalution}
In this section, we present an evaluation of our proposed framework. First, we describe the setup for our experiments in Section~\ref{sec:experimental setup}. Then, in Section~\ref{sec:experimentresults}, we evaluate the execution time of our framework to show its efficiency. Moreover,  we also present a range of experimental results for three representative CNNs to demonstrate that our novel framework can rapidly realize a wide variety of distributed CNN inference implementations with diverse trade-offs regarding energy consumption per device, memory usage per device, and overall system throughput. Finally, in Section~\ref{sec:comparison}, we analyze the effects on the energy consumption per device, the memory usage per device and the overall system throughput when scaling the distributed CNN inference to a varying number of deployed edge devices.
 
\subsection{Experimental Setup}
\label{sec:experimental setup}
% experimental setup
%the hardware platforms (i.e., NVIDIA Jetson Xavier NX embedded CPUs-GPUs MPSoc~\cite{jetsonnx}, gigabit switch) and the software methodologies (i.e., NSGA-II~\cite{nsga2} algorithm, DSE tool)
% introduce three networks and how to use NSGA-II to search mappings.
The goal of our experiments is to demonstrate that, thanks to our contributions presented in this paper, the \PM{} framework can easily and flexibly distribute CNNs over multiple edge devices. Moreover, it can do so with the same or higher CNN inference throughput, with lower per-device energy consumption, and with smaller per-device memory usage as compared to CNN execution on a single edge device. Since state-of-the-art CNNs have deep architectures with many layers, this leads to an immense variety of different CNN mappings on multiple edge devices, each having different characteristics in terms of energy consumption per device, CNN inference throughput, and memory usage per device. Therefore, we have designed a design-space exploration (DSE) experiment, using a Genetic Algorithm (GA), to find Pareto-optimal CNN mappings with respect to CNN inference throughput, energy consumption per device, and memory usage per device. 
%We introduce the experimental setup for this DSE experiment in three steps. First, we explain the three CNNs as well as the  hardware platforms used in our experiments. Second, we describe the evaluation process of the three different objectives for a given CNN mapping. Finally, we describe the actual DSE process, which is based on the well-known Non-dominated Sorting Genetic Algorithm (NSGA-II) ~\cite{nsga2} to generate different CNN mapping specifications and find the Pareto-optimal solutions. 

In our DSE experiment, we use three real-world CNNs, namely VGG-19~\cite{Simonyan15}, Resnet-101~\cite{He2015}, and Densenet-121~\cite{huang2019convolutional}, from the ONNX models zoo~\cite{onnxmodelzoo} that take images as an input for CNN inference. These CNNs are used in image classification and are diverse in terms of  types and number of layers, and memory requirements to store parameters (weights and biases). The first four columns in Table~\ref{CNNmodels+ET} list the details of the used CNN models. As these CNNs provide a good layer and parameter diversity, we believe that they are representative and good targets for our evaluations to demonstrate the merits of our framework. 

The aforementioned CNN models are mapped and executed on a set of up to eight edge devices where all devices are NVIDIA Jetson Xavier NX development boards~\cite{jetsonnx} connected over a Gigabit %unmanaged 
network switch. Each Jetson Xavier NX device has an embedded MPSoC  featuring six CPUs (6-core NVIDIA Carmel ARMv8) plus one Volta GPU  (384 NVIDIA CUDA cores and 48 Tensor cores). 
%We select NVIDIA Jetson Xavier NX as our experimental hardware platform because it is a well-known and easy-to-use embedded platform. 
%We can easily and accurately acquire the needed inference throughput and energy consumption data of each processor by setting timers within the executed code and by sampling the integrated onboard power monitors, respectively. 

For a given CNN mapping specification, we apply our \PM{} framework to generate and distribute a deployment package for every Jetson Xavier NX device. For every implementation generated by \PM{}, we measure and collect energy consumption per device, CNN inference throughput, and memory usage per device results, as an average value over 20 CNN inference executions. As the experiment is targeted to embedded devices, the batch size of CNN inference is 1. The inference throughput (measured by instrumenting the code with appropriate timers) and the memory usage per device are reported directly by the code itself during the CNN execution. To measure the energy consumption per device, a special sampling program reads power values from the integrated power monitors on each NVIDIA Jetson Xavier NX board during the CNN execution period, where the power consumption involves the whole board including CPUs, GPU, SoC, etc. 

To actually explore the different CNN mappings, while optimizing for the three target objectives (i.e., system throughput, energy consumption per device, and memory usage per device), we apply the well-known Non-dominated Sorting Genetic Algorithm (NSGA-II)~\cite{nsga2}. %NSGA-II is a well known, fast sorting and elite multi-objective genetic algorithm, widely used in many real-world applications, while today it can be considered as an outdated approach. But it is a quite useful algorithm to generate CNN mappings and select the elites in our experiments. 
%In this paper, we will only focus on the results found by NSGA-II. 
The chromosomes in our NSGA-II multi-objective GA implementation encode how a CNN is split into different segments and how these segments are mapped onto the various edge devices and resources within them. To evaluate the fitness of the encoded CNN mappings using our \PM{} framework, the chromosomes are translated to the framework's mapping format described in Section~\ref{sec:Interface}. In our DSE experiment, every CNN layer can be mapped either onto a single CPU core, onto six CPU cores, or onto a GPU inside an edge device. The GA is executed with a population size of 100 individuals, a mutation probability of 0.1, a crossover probability of 0.5, and performs 400 search generations. For all experiments with the three CNNs, the original data precision (i.e., float32) is utilized in order to preserve the original model accuracy of classification.

\subsection{Efficiency of \PM{} and DSE Results}
\label{sec:experimentresults}
%For each of the CNN models mentioned in Section~\ref{sec:experimental setup}, four independent DSE experiments are performed, targeting a configuration with single edge device, up to two devices, up to four devices, and up to eight devices, respectively. Here, every CNN layer can be mapped onto a single CPU core, six CPU cores, or a GPU inside an edge device. The exploration of the VGG-19 mappings on up to eight devices takes 7 days, whereas the DSE for the Densenet-121 and Resnet-101 mappings both take 5 days. Given that each GA-based DSE experiment evaluates 10,000 CNN mappings  (i.e., GA population size $\times$ GA iterations: 100 $\times$ 100 = 10,000) using our \PM{} framework, this means that each mapping evaluation only takes about one minute or less. Here, an evaluation involves running \PM{} to split a CNN, generate code for and deploy packages to the involved edge devices, and actually executing the CNN (for 20 times, see Section~\ref{sec:experimental setup}) in a distributed fashion.
We start with evaluating the execution time of \PM{} itself, to provide insight on how long the framework generally takes to split a CNN model (front-end), to generate the code for the distributed CNN execution (back-end), and to deploy the generated packages to the edge devices for actual execution. To this end, we have measured the required time for each of these phases using the 'worst-case scenario' in the scope of our experiments: using the maximum number of splits in our CNNs to generate sub-models (24 splits/sub-models of a CNN in our experiments), and mapping and deploying the generated sub-models to the maximum number of edge devices (8 in our experiments). These measurements were done on
a system equipped with an Intel Core i7-9850H processor, running Ubuntu 20.04.3 LTS. The last three columns in Table~\ref{CNNmodels+ET} provide a breakdown of the execution time (in seconds) of \PM{} for the three CNNs in these worst-case scenarios. 
From the results in Table~\ref{CNNmodels+ET}, we can see that \PM{} is able to produce executable, distributed CNNs and deploy them on the various edge devices in a relatively short time frame, i.e., in less than a minute for any of the three used CNNs in our worst-case scenario. The comparatively larger execution time of the front-end for VGG-19 is due to the high number of parameters in this model, and the resulting overheads in \PM{} of copying these parameters to the large number of sub-models. In any case, these results demonstrate that \PM{} allows for rapidly splitting CNNs and deploying them for distributed execution on multiple edge devices.

\begin{figure*}[!t]
    \centering
    \begin{subfigure}[t]{0.32\linewidth} 
    \includegraphics[width=\linewidth]{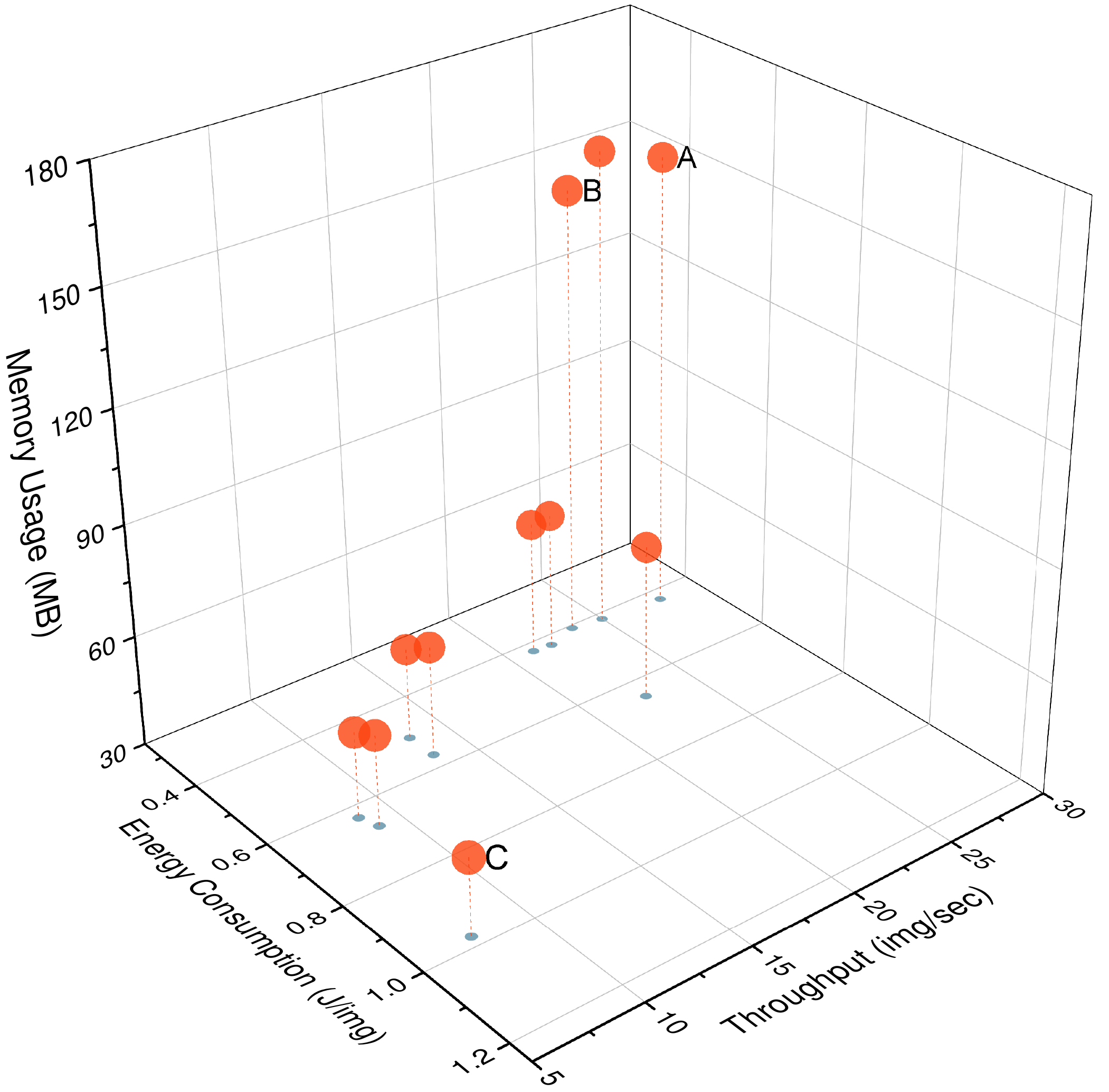}
        \centering
         \caption{DenseNet-121 (910 layers)}
         \label{densenet121}
    \end{subfigure} % <-- added
    \begin{subfigure}[t]{0.32\linewidth} 
        \includegraphics[width=\linewidth]{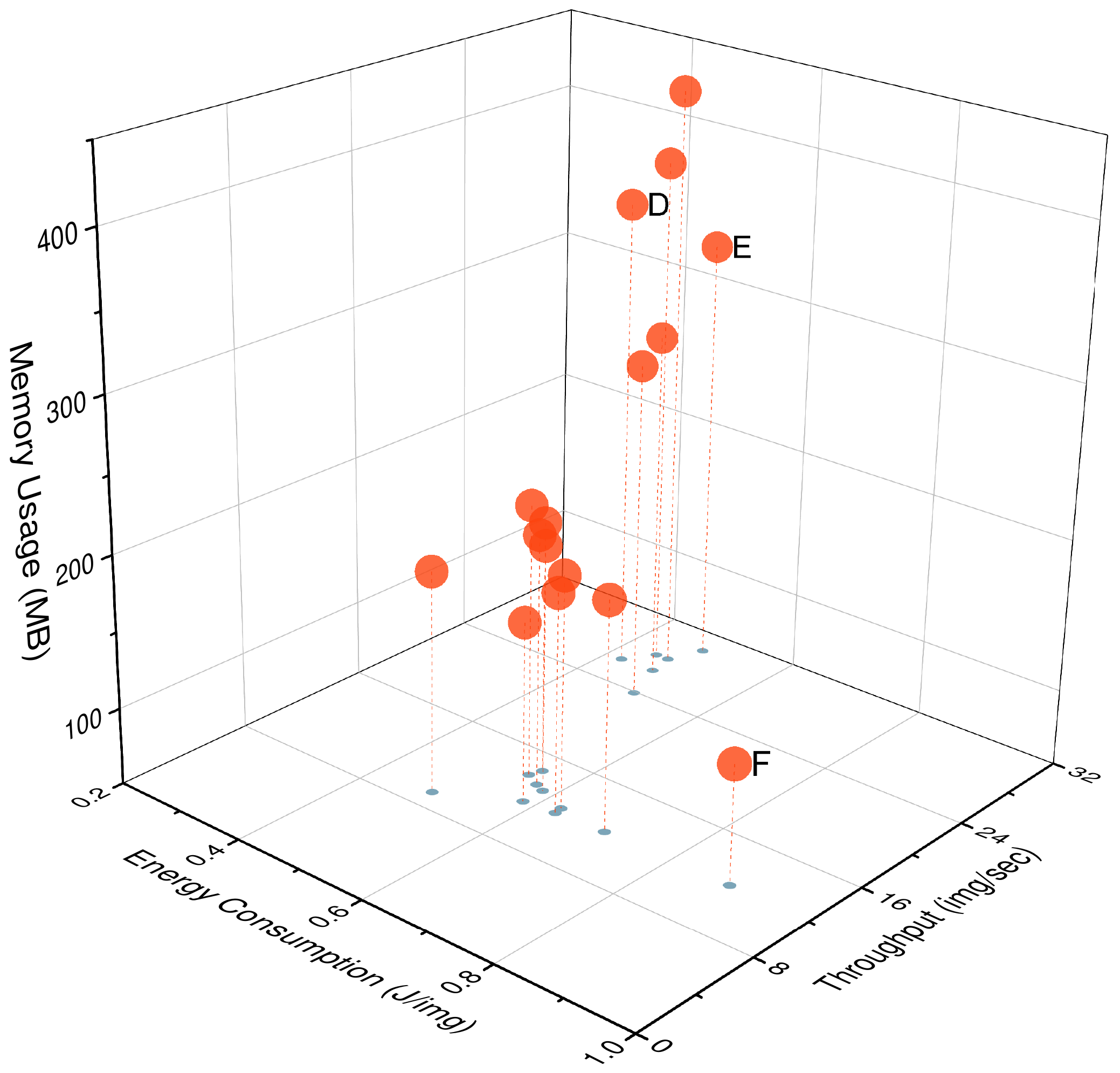}
        \centering
        \caption{ResNet-101 (344 layers)}
        \label{resnet101}
    \end{subfigure} % <-- added
    \begin{subfigure}[t]{0.32\linewidth}  
    \includegraphics[width=\linewidth]{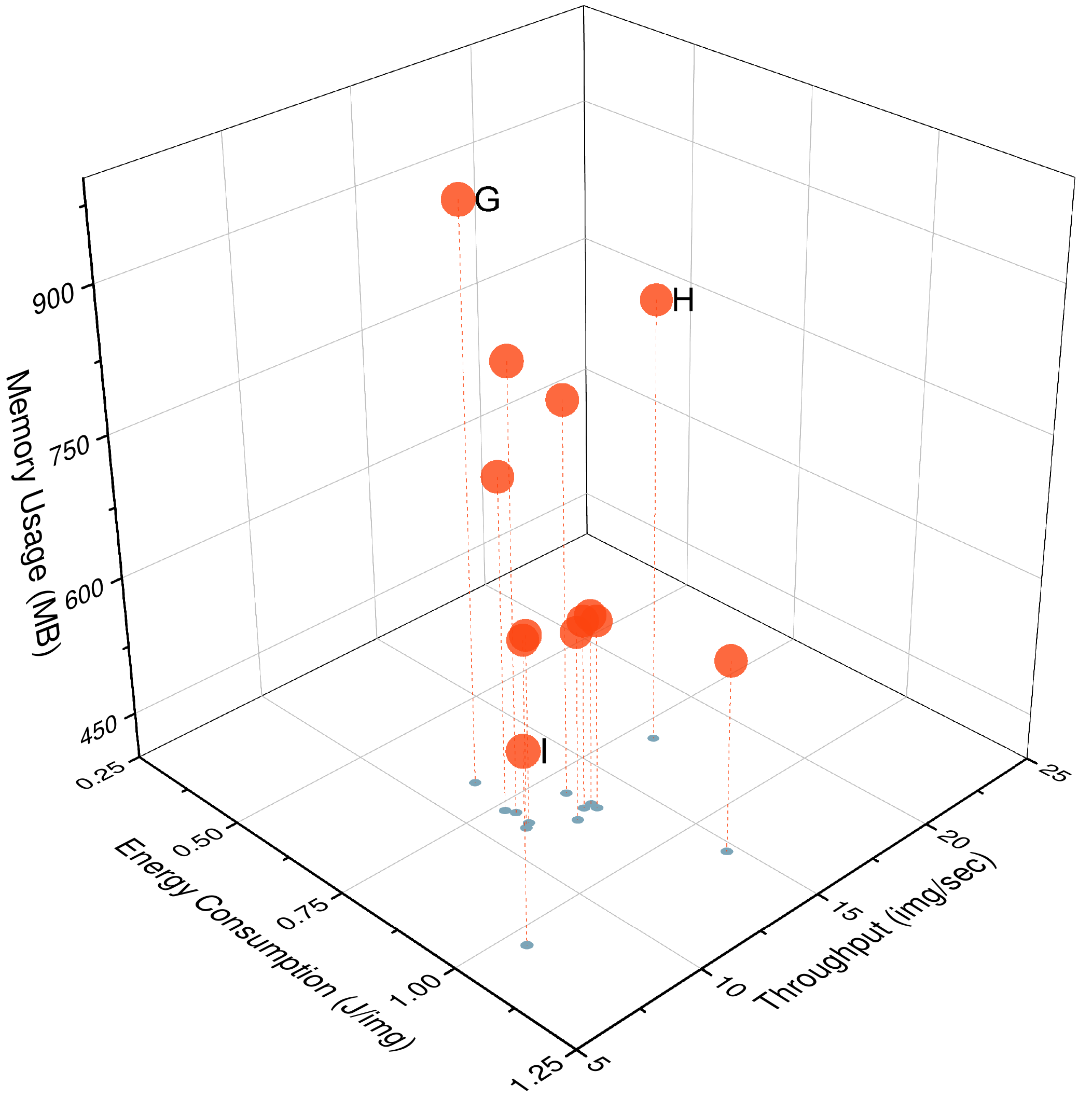}
            \centering
      \caption{VGG-19 (47 layers)}
      \label{vgg19}
    \end{subfigure}
\caption{Pareto-optimal CNN mappings from our DSE experiment with three CNNs.} 
\label{paretopoints}
\end{figure*}

Our DSE experiment explores a wide range of different CNN mappings and results in a Pareto front with several Pareto-optimal mappings. In such a set of Pareto-optimal mappings, none of the targeted objectives (energy consumption, throughput, and memory usage) can be further improved without worsening some of the other objectives. More specifically, we consider the \textit{maximum} energy consumption \textit{per device}, \textit{maximum} memory usage \textit{per device}, and total system (CNN inference) throughput as our target objectives. Figures~\ref{densenet121},~\ref{resnet101}, and ~\ref{vgg19} show the Pareto-optimal CNN mappings found by our DSE for DenseNet-121, ResNet-101, and VGG-19, respectively. 
%These Pareto-optimal mappings are gathered  from the four aforementioned independent DSE experiments with configurations of up to one device, two devices, four devices and eight devices.  
%For each of the Pareto-optimal mappings (orange dots in the graphs), a projection (small dots) is also shown on the three planes spanned by the three target objectives.  
To better illustrate (the diversity of) these Pareto-optimal mappings, Table~\ref{tablepoints} shows more details about a selection of these mappings (points A to I in Figure~\ref{paretopoints}) for comparison. As a reference, the table also includes the mapping results when using a single edge device with 6 CPUs or 1 GPU.
%The first column in the table lists the three different CNNs, followed by the second column with the selected Pareto points from the Figure~\ref{paretopoints}. We also include the evaluation results of single device with all six CPU cores or with the GPU to show comparison with these selected points.
Columns 3 and 5 show the maximum energy consumption per device and maximum memory usage per device for a specific CNN mapping, respectively. Column 4 shows the overall system throughput. Columns 6, 7 and 8 show the hardware configurations of the selected CNN mappings,  consisting of the number of deployed edge devices, the total number of used CPU cores and the total number of used GPUs, respectively. 

\begin{table*}[t]
  \caption{Selected Pareto-optimal Mappings (points) from Figure~\ref{paretopoints}}
  \label{tablepoints}
  \centering
\begin{tabular}{@{}lccccccc@{{}}}
\toprule
Network & Points & Max. per-device & System & Max. per-device & \# Edge Devices & \# CPU cores & \# GPUs \\ 
 &  & Energy (J) & Throughput (FPS) & Memory (MB) &  &  &  \\ 
\midrule
\midrule
 & 1-Device CPU      & 0.905       & 7.987       & 129.984         & 1           & 6        & 0   \\ 
  & 1-Device GPU      & 0.650      & 12.807       & 251.172        & 1           & 0        & 1   \\  
 DenseNet-121 & A      &  	0.430		     & \textbf{27.941}       &  152.336	      & 4           & 0        & 4   \\
 & B  &     \textbf{0.408}		     & 23.551             & 149.941           & 6           & 6       & 5   \\   
 & C      & 0.977	 	    &7.546       & \textbf{51.066}         & 8           & 38        & 0   \\

%\bottomrule
\midrule
\midrule
 & 1-Device CPU      & 1.635       & 5.786      & 656.527         & 1           & 6        & 0   \\     
  & 1-Device GPU      & 1.031       & 21.767       & 955.012         & 1           & 0        & 1   \\ 
 ResNet-101 & D      &  \textbf{0.425}	  &  	26.406  &  	360.766         & 7           & 0        & 7   \\
  & E      & 0.488	 & 	\textbf{30.048}      & 329.641           & 7           & 12        & 5   \\
 & F      & 0.886			& 12.123	     & \textbf{127.883}         & 8           & 48       & 0   \\ 
%\bottomrule
\midrule
\midrule
 & 1-Device CPU      & 1.471      & 7.273       & 1310.91         & 1           & 6        & 0   \\
  & 1-Device GPU      & 1.523       & 11.664       & 1666.418         & 1           & 0        & 1   \\
 VGG-19 & G      &   \textbf{0.680} 			   & 11.651       &	998.273        & 6           & 0        & 6   \\
 & H      &   	0.791        & 	\textbf{17.385}         & 	868.496	          & 6           & 6       & 5   \\ 
 & I      &   		1.035	        & 7.194          & \textbf{604.504}          & 7           & 30      & 2   \\  

\bottomrule
\end{tabular}
\end{table*}

From Figure~\ref{paretopoints} and Table~\ref{tablepoints}, we can see that \PM{} allows for easily and rapidly realizing a wide variety of distributed CNN inference implementations with diverse trade-offs regarding per-device energy consumption, per-device memory usage, and overall system throughput.
%A number of observations can also be made from the obtained Pareto-optimal mappings. For example, the maximum energy consumption and memory usage per device typically reduces (i.e., due to less computations and parameters per device) while the system throughput often improves (due to exploited pipeline parallelism) when the number of deployed edge devices increases. 
Taking point A as an example, a distributed execution of DenseNet-121 on four devices utilizing only GPUs can reduce the maximum energy consumption per device by 52.5\% and 33.8\% as compared to the 1-Device CPU and 1-Device GPU hardware configurations, respectively. The system throughput of DenseNet-121 on four devices achieves a 3.5x and 2.2x performance improvement compared to the 1-Device CPU and 1-Device GPU configurations, respectively.
In terms of per-device memory usage, the CNN mapping A with four devices consumes 39.3\% less memory than the 1-Device GPU implementation, but consumes 17.2\% more memory as compared to the 1-Device CPU configuration.
 
An observation that can be made in general from our DSE results is that by increasing the number of utilized devices, the per-device memory usage is not always reduced if GPUs are deployed within (some of) the devices. In Table~\ref{tablepoints}, this is clearly illustrated by, for example, CNN mappings A and B. These mappings have even higher per-device memory usage when distributing the CNN over, respectively, four and six devices as compared to a 1-Device CPU configuration. The higher memory usage when deploying GPUs is due to the fact that an NVIDIA Jetson Xavier NX device has 8GB memory that is shared between CPU and GPU programs. During the loading phase of CNN models, there will typically be at least two copies of the CNN weights when using the GPU: those from the original model file in the host memory, and those initialized as part of the GPU engine.

%Evidently, the maximum energy consumption and system throughput are strongly correlated: A higher system throughput implies that distributed devices take a shorter time both for computation and communication, which typically means that the energy consumption per board also becomes smaller. In contrast, the number of GPUs and memory usage are negatively correlated. A NVIDIA Jetson Xavier NX device has 8GB memory which is shared between CPU and GPU programs. During the loading phase of CNN models, there will typically be at least two copies of the CNN weights: those from the original model file in the host memory, and those initialized as part of GPU engine. Thus, the design points with lowest memory consumption usually do not use GPUs for computation. 

\subsection{Varying the Number of Edge Devices}
\label{sec:comparison}
\begin{figure*}[!t]
\centering
  \includegraphics[width=\textwidth]{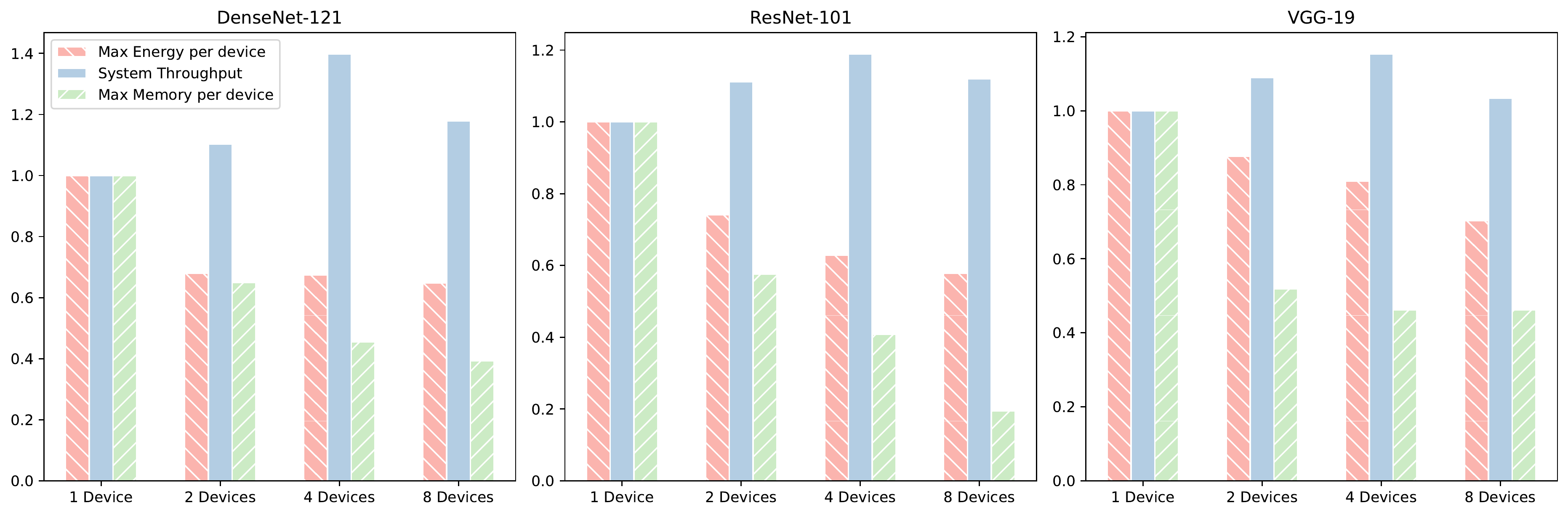}
  \caption{System throughput and max energy/memory per device when varying the number of edge devices for three CNNs.}
  \label{comparison4dse}
\end{figure*}

In Figure~\ref{comparison4dse}, we show the effects on the maximum per-device energy consumption, maximum per-device memory usage, and system throughput when scaling the number of deployed edge devices in the distributed CNN execution. Every bar in Figure~\ref{comparison4dse} reflects the best value (energy consumption, memory usage, or throughput) found among all the evaluated mappings, during our DSE experiment, with a specific number of deployed edge devices. This implies that the value reflected by each bar may come from a different Pareto-optimal mapping. For better visualization, all results in Figure~\ref{comparison4dse} have been normalized, where the results for a configuration with one edge device are taken as the reference (i.e., these represent the results of the best found mappings when targeting a single edge device).

From Figure~\ref{comparison4dse}, we can see that, in general, both the per-device energy consumption and the per-device memory usage can be improved (i.e., reduced) when increasing the number of deployed edge devices. Evidently, this is due to the fact that the workload (the size and/or the number of executed sub-models) on each participating edge device is reduced when increasing the number of edge devices. Moreover, in some cases, the improvement can be significant. For example, for ResNet-101, the maximum per-device energy consumption and maximum memory usage are reduced by around 40\% and 80\%, respectively, when distributing the CNN over eight edge devices as compared to execution on a single device. 
Furthermore, the results in Figure~\ref{comparison4dse} show that the system (CNN inference) throughput can also be improved by means of distributed CNN execution. This is because of the exploitation of pipeline parallelism in the distributed CNN execution. For example, for DenseNet-121, ResNet-101, and VGG-19, the inference throughput increases by up to 38\%, 18\%, and 18\%, respectively when executing the CNN inference on up to four edge devices as compared to a single device. However, the inter-device data communication overheads involved in distributed CNN execution may prevent any further throughput gains, or even cause a slowdown, when scaling the CNN execution to a larger number of edge devices. For example, for all three CNNs, DenseNet-121, ResNet-101, and VGG-19, we see a slowdown in system throughput when scaling the CNN inference from four to eight edge devices.

\section{Discussion}
\label{sec:Discussion}
Our current \PM{} framework implementation seeks to provide the greatest flexibility in terms of facilitating distributed execution of CNN models on a wide range of different hardware configurations at the Edge, i.e., configurations different in the number of deployed edge devices as well as in the nature (architecture) of these devices. Therefore,  in the current version of \PM{}, we have integrated our own custom CNN Inference Library (based on the NCNN~\cite{ncnn} and Darknet~\cite{darknet13} frameworks) that supports CNN implementation and execution on a variety of hardware platforms (e.g., Raspberry Pi, NVIDIA Jetson, etc.). Our own custom library is not optimized for specific devices in order to provide the greatest possible flexibility. With our focus on flexibility, we have not yet heavily invested in the performance optimization of our \PM{} framework when, e.g., targeting specific edge devices. For example, in the future, we plan to integrate the TensorRT framework into \PM{} to support very optimized and efficient CNN execution when targeting specific NVIDIA-based devices such as the NVIDIA Jetson series of embedded computing boards because TensorRT has demonstrated to produce superior CNN inference performance on NVIDIA-based devices \cite{Ulker2020}. 

%Though the results obtained by using our framework are comparable with TensorRT now, it's not feasible to compare with TensorRT for several reasons. The TensorRT engine takes a very long time to serialize and also consumes a very large static memory to achieve acceleration, which leads to its energy consumption per device and maximum memory usage per device extremely high.

\section{Conclusions}
\label{sec:conclusion}
In this paper, we have presented \PM{}, the first fully automated framework for distributed CNN inference over multiple resource-constrained devices at the Edge.  The framework features a unified and flexible user interface, fast CNN model partitioning and code generation, and easy deployment of the CNN partitions on edge devices. We have demonstrated the flexibility of \PM{} with a detailed example illustrating all main steps in the \PM{} design flow. By applying the design flow on three representative CNNs, we have evaluated \PM{} in terms of  efficiency and usefulness in facilitating fast and accurate Design Space Exploration (DSE). Our DSE experiments and results show that \PM{} can easily and rapidly realize a wide variety of distributed CNN inference implementations on multiple edge devices, achieving improved (i.e., reduced) per-device energy consumption and per-device memory usage as well as improved system (inference) throughput. It is worth noting that these improvements are achieved without losing the initial CNN model accuracy because the steps in our framework change neither the CNN layers and their data dependencies nor the values and precision of the CNN parameters (weights and biases).    

\balance
\bibliography{IEEEabrv,./bib/fl.bib, ./bib/edgecloud.bib, ./bib/edge.bib}

\end{document}